\IfFileExists{submit.txt}{
  \documentclass[12pt]{aastex}   %,preprint
}{
  \documentclass{emulateapj}   %,preprint
} 
\usepackage{natbib,color}

\newcommand{\p}[1]{{#1}}
\newcommand{\pc}[1]{{#1}}

\newcommand{\rmd}{{\mathrm d}}
\renewcommand{\vec}[1]{ {\mathbf #1} }

\newcommand{\B}{\vec B}
\newcommand{\BI}{{\vec B}_{\rm I}}
\newcommand{\Bp}{{\vec B}_{\rm p}}
\newcommand{\Bpx}{B_{\rm p,x}}
\newcommand{\hcrit}{h_{\rm crit.}}
\newcommand{\li}{l_{\rm i}}
\newcommand{\lf}{l_{\rm f}}
\newcommand{\Lc}{L_{\rm c}}
\newcommand{\Ls}{L_{\rm s}}
\newcommand{\Le}{L_{\rm e}}
\IfFileExists{submit.txt}{
  \renewcommand{\na}{n_{\rm a}}
}{
  \newcommand{\na}{n_{\rm a}}
} 
\newcommand{\nBp}{n_{\rm Bp}}
\newcommand{\nBpcrit}{n_{\rm Bp, crit.}}
\newcommand{\nI}{n_{\rm I}}
\newcommand{\nr}{n_{\rm r}}
\newcommand{\pta}{point ``a''}
\newcommand{\ptap}{point ``a$^{\prime}$\,''}
\newcommand{\ptb}{point ``b''}
\newcommand{\ptc}{point ``c''}
\newcommand{\ptd}{point ``d''}
\newcommand{\rep}{r}
\newcommand{\sh}{S_{h}}
\newcommand{\uf}{u}

\begin{document}
\title{Criteria for Flux Rope Eruption: Non Equilibrium versus Torus Instability}
%\subtitle{Non equilibrium versus torus instability}
\author{P. D\'{e}moulin and G. Aulanier} 
\affil{LESIA, Observatoire de Paris, CNRS, 5 place Jules Janssen, 92190 Meudon, France} 
\email{Pascal.Demoulin@obspm.fr}

\begin{abstract}
   % context
The coronal magnetic configuration of an active region typically evolves quietly
during few days before becoming suddenly eruptive and launching a coronal mass ejection (CME).  The precise origin of the eruption is still debated.  Among several mechanisms, it has been proposed that a loss of equilibrium, or an ideal magneto-hydrodynamic (MHD) instability such as
the torus instability, could be responsible for the sudden eruptivity. Distinct approaches have also been formulated for limit cases having circular or translation symmetry.     % aims, methods
We revisit the previous theoretical approaches, setting them in the same analytical framework.  The coronal field results from the contribution of a non-neutralized current channel added to a \pc{background magnetic field, which in our model is the potential field generated by} two photospheric flux concentrations.
The evolution on short Alfv\'enic time scale \p{is governed} by ideal MHD. 
    % results 
We show analytically first that the loss of equilibrium and the stability analysis are
two different views of the same physical mechanism.  Second, we identify that the same physics is involved in the instability of circular and straight current channels. Indeed, they are just two particular limiting case of more general current paths. 
  %We compare the instability thresholds in the limit of straight and circular current channels, finding that they are closely comparable for thick current channels (as present in the MHD simulation and as expected in the corona) while these thresholds are well distinct at the limit of very thin current channels (as typically found in previous studies). 
A global instability of the magnetic configuration is present when the current channel  
is located at a coronal height, $h$, large enough so that the decay index of the potential field, $\partial \ln |\Bp|/\partial \ln h$ is larger than a critical value.
At the limit of very thin current channels, previous analysis found a critical decay index of $1.5$ and $1$ for circular and straight current channels, respectively.
However, with current channels being deformable and as thick as expected in the corona, we show that this critical index has similar values for circular and straight current channels,
typically in the range [1.1,1.3]. 
% This is significantly below the critical indices found in MHD simulations (in the range [1.5,1.9]), indicating that further theoretical developments are required.
 
\end{abstract}


\section{Introduction}
  \label{Introduction}

  % {\S\bf Generality on CMEs. }\\ 
   A coronal mass ejections (CME) is the consequence of the sudden destabilization of a part of the coronal magnetic field.  The eruption is preceded by a long phase (days to week) during which the magnetic field is progressively stressed and free magnetic energy builds up.  The configuration typically grows quasi-statically (with velocities well below the Alfv\'en velocity).  At a point of the evolution, in a few minutes up to an hour, the system becomes very dynamic, with a global upward motion, as traced by the evolution of the cold \p{plasma in the associated filament and of the hot plasma in coronal loops.}  Later on, a significant release of magnetic energy occurs, and a flare is typically observed.  If the downward magnetic tension of the covering magnetic arcade is weak enough, the erupting plasma and magnetic field is launched towards the interplanetary space as a CME.   In summary, the CME phenomena occurs in four main phases: build-up, instability, acceleration, and propagation.  They have been reviewed by \citet{Forbes06} and \citet{Vrsnak08}.
   
  % {\S\bf What is known in the four phases.}\\ 
The two last phases are the most spectacular ones, so they are better constrained
by observations and they are more deeply modeled, in particular with MHD simulations \citep[e.g.][]{Amari03b,Fan07,Torok07}.  The first phase is a slow evolution and it is usually difficult to characterize in observations what are the generic key points which lead to eruption.  The main physics which emerges from observations is the presence of new emerging magnetic flux, progressive dispersion of the whole flux, the build up of a very sheared field in the vicinity of the photospheric inversion line (PIL), and the cancelation of flux at the PIL \citep[e.g.][]{vanDriel98,Green02,vanDriel03}.   At the least the three last are physically related since flux dispersion lead to the convergent flows towards the PIL, increasing the magnetic shear and forcing flux cancelation.  This also implies the build-up of a flux rope with J-shaped coronal loops transformed by reconnection into S-shaped loops \citep[e.g.][]{Moore95,Gibson06b,Green09}.

  % {\S\bf Importance to understand why it gets unstable}\\ 
A still open issue is why does the magnetic configuration erupt? There is usually no evidence of a large amount of new magnetic flux (with a magnitude comparable to the pre-eruptive flux), so the eruption is not driven by the sub-photospheric evolution but rather the coronal magnetic configuration \p{becomes
unstable} at some point during the slow evolution.  During this phase, magnetic reconnection is probably involved as a key mechanism for the progressive transformation of the magnetic configuration. However,
an ideal instability is thought to initiate the CME since the upward acceleration phase starts before the impulsive phase of the flare in the majority of events \citep[e.g.][]{Kahler88,Maricic07}.  Later on, magnetic reconnection plays a key role in the eruption as the peak of the upward acceleration is typically found correlated with the peak of the hard X-rays and of the time derivative of soft X-rays flux \citep[e.g.][]{Neupert01,Zhang01,Vrsnak04b,Temmer08}. 

  % {\S\bf First phase}\\ 
\p{Magnetic reconnection also plays a key role during the first
phase as it permits the progressive transformation of very sheared field lines into
a twisted flux rope.}  However, the MHD simulations of \citet{Aulanier10} have shown that magnetic reconnection at the photospheric level, or later on in the corona below the flux rope, is not directly responsible for the onset of the eruption.  The configuration rather gets unstable when the flux rope reaches a height where the potential field, associated to the distribution of the photospheric magnetic flux, is decreasing fast enough with height.  \p{This relates the
onset of the eruption in MHD simulations to a series of analytical studies, as
summarized below.}

%   {\S\bf Evolution of erupting fields}\\ 
%\citep{Chen03,Chen07,Vrsnak07,Schrijver08}

%%%%%%%%%%%%%%%%%%%%%%%%%%%%%%%%%%%%%%%%%%%%%%%%%%%%%%%%%%%%%%%%%%%%%%%%%%
\begin{figure}[t!]    
\IfFileExists{submit.txt}{
  \centerline{\includegraphics[width=0.5\textwidth, clip=]{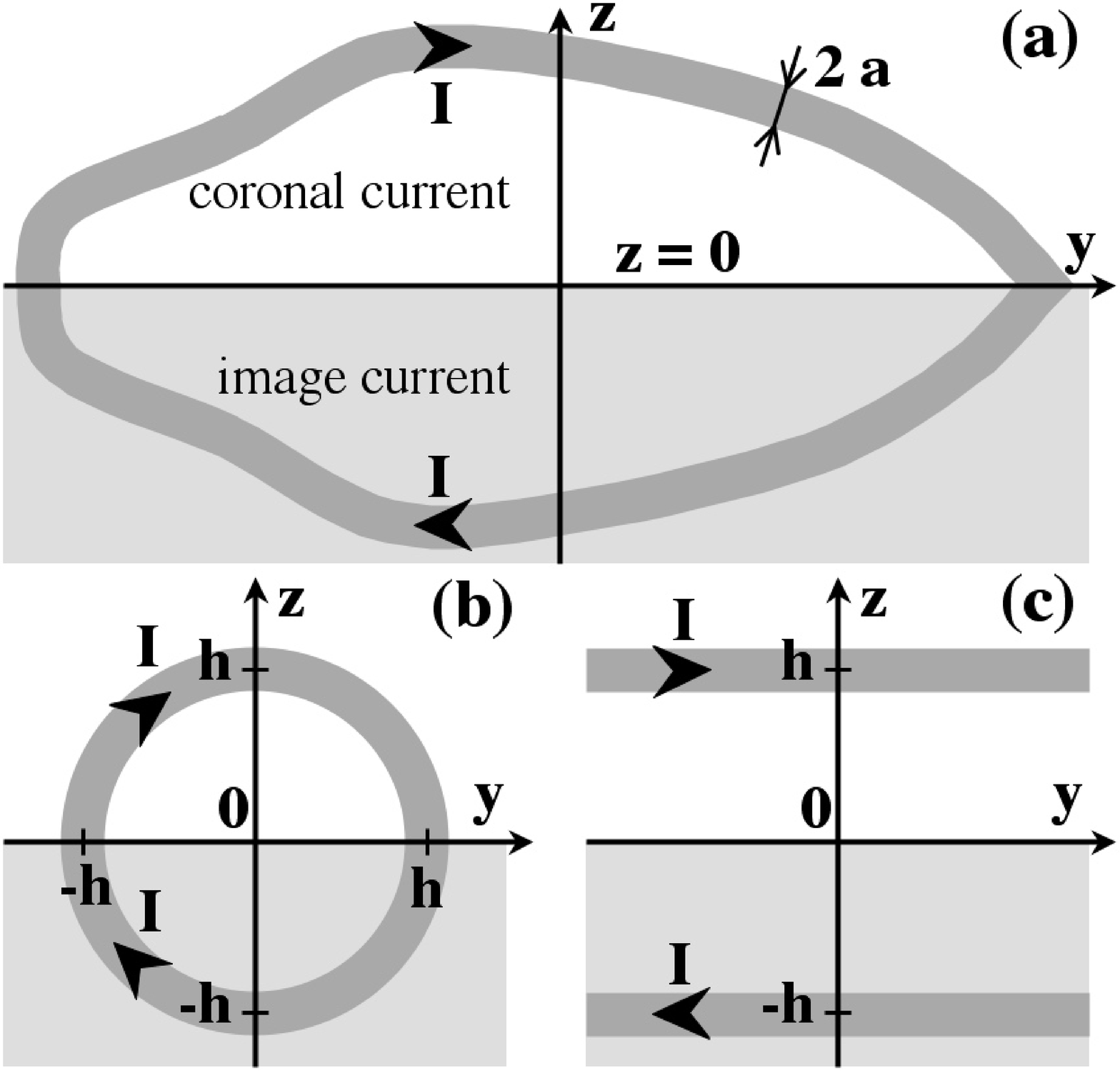}}
}{
  \centerline{\includegraphics[width=0.4\textwidth, clip=]{f1}}
} 
\caption{{\bf (a)} Schema of a current channel in the corona and its image current below the photosphere (located at $z=0$). The current channel has a radius $a$.  
        {\bf (b,c)} Particular cases with a circular and straight current channels, respectively.
}
 \label{Fig_schema}
\end{figure}

  % {\S\bf Introduction of image current}\\ 
  The equilibrium of a flux rope within a coronal field was first considered by 
\citet{Kuperus74} with the following simplifications.  In Cartesian coordinates, they modeled a flux rope with the magnetic field created by a straight line current of intensity $I$, located at a height, $z=h$, above the photosphere located at $z=0$ (Figure~\ref{Fig_schema}(c)).  The flux rope field is added to \pc{a simple background magnetic field: a} potential field, $\Bp$, associated to a bipolar photospheric magnetic field.     They included the observed \p{insignificant} evolution of the vertical component, $B_z$, at the photospheric level during the instability phase of a CME by introducing an image current of intensity $-I$, located at the height $z=-h$.  The physical result is that two oppositely directed Laplace forces are acting on the coronal line current: one from the potential field, $\Bp$, and the other from the magnetic field created from the image current (or the equivalent surface current at $z=0$).  The equilibrium is then described by a curve $I(h)$.
   
  % {\S\bf 2D straight currents. circuit models }\\    
\citet{vanTend78} showed that, $I(h)$ is an increasing function of $h$ at low height and that it has a local maximum if the horizontal component of $\Bp$ orthogonal to the line current, $|\Bpx|$, decreases fast enough with height.   Then, supposing that the current $I$ can be increased progressively to larger values during the buildup phase, a loss of equilibrium occurs at $h=\hcrit$ defined by the maximum of the function $I(h)$.   This occurs where $|\Bpx |$ decreases faster than $1/h$.  This model was later developed within a circuit theory, introducing an electric potential and a resistance in the circuit to describe the temporal evolution of $I$.  The current sheet which forms below the erupting flux rope was described by another electric circuit.  This provides a set of coupled equations which describes the main phases of an eruptive flare \citep{Martens89,vanBallegooijen89}.  Next, the \pc{background} potential field, $\Bp$, was replaced by a linear force-free field to include, as observed, a sheared field \citep{Amari89}.  A loss of of equilibrium is also present if $|\Bpx |$ is decreasing fast enough with height \citep{Demoulin88}.

  % {\S\bf Ideal MHD constraint: flux conservation}\\
In the above theory, it is supposed that the current intensity, $I$, is the main driver of the evolution and that it can be increased to arbitrary large values.  However, the coronal physics is not precisely described by any circuit theory because the electric current is a consequence of the force balance together with MHD constraints \citep[e.g.][]{Parker96}. \p{The effect of magnetic reconnection on the system (through changing field line connectivity and energy release) before the flare/CME acceleration phase is minimal and therefore the stability of an equilibrium is typically tested in ideal
MHD. }
The conservation of the coronal magnetic flux passing below the flux rope is typically used to set a constraint on the current evolution. \citet{Anzer90} claimed that there is no longer a loss of equilibrium with this ideal MHD constraint, but \citet{Demoulin91} found that this constraint mostly displaced the loss of equilibrium point to a larger height (slightly after the maximum of the $I(h)$ curve), in agreement with the MHD simulation of \citet{Forbes00}.  The following developments have shown that a loss of equilibrium is typically present if the flux rope radius is thin enough. It occurs, for example, when the photospheric polarities are subject to converging motions toward the PIL, or when their magnetic flux is decreased,  even when ideal MHD is assumed during the full build-up phase, \citep{Isenberg93,Forbes91,Forbes95}.

  % {\S\bf Introduction of curvature. Loss of equilibrium}\\ 
When the current channel is curved, an extra force is present, called the hoop force \citep[e.g.][]{Bateman78}.  The electric current of a curved channel creates a magnetic field component orthogonal to the channel. This implies an outward Laplace force (away from the curvature center).   In terms of the magnetic field, this force is due to the over magnetic pressure of the azimuthal field component present in the direction of the curvature center.  This force is at the heart of the magnetically driven model of \citet{Chen89} and subsequent developments \citep[e.g.][]{Garren94,Krall00}.  This force is also present in MHD models where the straight line current of previous paragraph is replaced by a ring of current \citep{Lin98,Lin02,Titov99}.   As for the above cartesian models, an ideal MHD evolution during the build-up phase also typically, but not always, lead to a loss of equilibrium.  The differences will be discussed in Sections~\ref{Loss} and \ref{Stab}.   A non equilibrium point could also be present when line-tied conditions are imposed at the photospheric footpoints of the flux rope \citep{Isenberg07}.

  % {\S\bf Instability with curved current channel}\\ 
From another point of view, \citet{Kliem06} studied the stability of a toroidal current ring immerged in a background potential field, $\Bp$.   Extending the results summarized in \citet{Bateman78}, they derive that an instability occur when the background field component orthogonal to the torus ($|\Bpx |$) decreases faster than $1/h^{3/2}$ with a correction depending on the torus aspect ratio (major over minor radius).   \citet{Kliem06} called it the ``torus instability''.   They analyzed cases where the electric current $I$ was held constant or fixed by the conservation of the total magnetic flux within the torus hole.  
 
  % {\S\bf Roadmap of the paper}\\ 
The aims of present paper is to revisit the above studies to analyze their \p{relationships}.
 Is this ``torus instability'' different from the loss of equilibrium found in previous studies with a toroidal current?  Has it a different physical origin than the instability of \citet{vanTend78} obtained with a straight current channel? 
The Cartesian and axisymmetric models are first written with the same formalism in Section~\ref{Basic}.  We compare the loss of equilibrium and the instability approaches in both geometries in Section~\ref{Loss}, before comparing the criteria of instabilities taking into account a finite current-channel width which evolves during the perturbation (Section~\ref{Stab}).   \p{Finally,
in Section~\ref{Conclusion}, we summarize our results and discuss the relationship between the
Cartesian and axisymmetric models, as well as} between the loss of equilibrium and the torus instability.

%%%%%%%%%%%%%%%%%%%%%%%%%%%%%%%%%%%%%%%%%%%%%%%%%%%%%%%%%%%%%%%%%%%%%%%%%%%
\section{Basic Concepts}
  \label{Basic}

   The coronal magnetic field $\B$ can always be is written as the sum of 
\pc{a background magnetic field $\Bp$ and the magnetic field $\BI$, created by localized coronal currents and their images below the photosphere (Section~\ref{Basic.2}).  
$\BI$ has a vanishing $B_z$ component at $z=0$ by construction.  All the analytical models use this field decomposition with various approximations for $\BI$.  
$\Bp$ is most frequently taken as the unique potential field associated with the photospheric ``normal magnetogram'' $B_z(x,y,0)$ and with vanishing strength at infinite distance. We consider this case below, however, it is worth noting that, in general, $\Bp$ can also incorporate distributed coronal currents \citep[e.g. $\Bp$ could be a linear force-free field, and $\BI$ needs to be self consistently computed,][]{Demoulin88}.}  

\subsection{Modelization with Concentrated Currents} %%%%%%%%%%%%
  \label{Basic.1}

   %{\S\bf Thickness}\\ 
In order to have a set of equations solvable analytically, simplifications in the magnetic configuration \p{need to be made}.  A first one is to suppose that the electric current is restricted to a thin channel in the corona (Figure~\ref{Fig_schema} (a)).  More precisely that its typical thickness, $2a$, is small compared to the spatial scales of $\Bp$ and to the local radius of curvature of the current channel axis.   In an active region, the magnetic shear is typically concentrated around the PIL, while the surrounding arcade is more potential, so electric current are stronger around and above the PIL.  With reconnection of sheared loops at the PIL, an important fraction of the currents are inside a twisted flux tube \citep[][and references therein]{Aulanier10}. Then, the introduction of a concentrated current channel is motivated by observations and MHD simulations, however it is still an important simplification (e.g., neglecting the effect of more distributed currents as well a the presence of narrow current layers and sheets).

   %{\S\bf Internal equilibrium}\\
The \p{approximation} of a thin current channel allows to separate approximately the magnetic equilibrium in a internal and external equilibrium \citep{Chen89,Isenberg93}.  This splitting is better achieved as the current channel is thinner.  For the external equilibrium, the Laplace force, integrated over the channel cross section, vanishes, so that there is no average magnetic field component orthogonal to the current channel.  The internal equilibrium is solved locally, in the channel cross section.
   %, since there is no average external magnetic field.  
A twisted flux tube has generically both toroidal (axial) and poloidal (azimuthal) magnetic field and electric current components.
The equilibrium is typically solved in cylindrical coordinates with a balance between the total magnetic pressure gradient and the tension of the poloidal magnetic field (force-free field solution).   
This internal equilibrium is not the object of present paper, and we refer to the work of \citet{Lin98}. 

   %{\S\bf Return current}\\ 
Another important simplification is the absence of a neutralization, or return, current around the direct current.  Such return current, of opposite direction and with the same magnitude as the direct current, is expected to be present in a magnetic field formed by emergence or induced by localized boundary (photospheric) motions \citep{Parker96b}. In both cases the complete neutralization is due to a vanishing circulation of $\vec{B}$ around a large path enclosing the current channel.  Return currents are indeed present in MHD simulations, but when a significant magnetic field component is present along the PIL, the direct current has a larger magnitude than the return current \citep{Torok03,Aulanier05}.  Indeed, only a partial neutralization was typically reported in sunspots \citep{Wheatland00,Venkatakrishnan09}.  With partial neutralization, the current intensity, $I$, is the non-vanishing sum of the two opposite currents.     

\pc{
We argue that the occurence of non fully neutralized currents should be a common 
feature in solar active regions. The contrary would imply that current-carrying 
flux tubes should be fully surrounded by potential fields, not only high up in 
the corona, but also low down around PILs. It is worth noticing, however, that 
some MHD models for solar eruptions clearly do not require a net current \citep[the 
magnetic breakout, tether cutting and flux disappearance models, respectively 
addressed by][]{Antiochos99,Moore92,Amari00}, as they are based on a gradual diminishing of the tension of the background field, irrespectively of the distribution of electric currents at lower altitudes. Many other models exist \citep[as reviewed e.g. 
in][]{Forbes06,Aulanier10}, but it is conceivable that several actually 
fall into the physical frame studied in the present paper. 
}

\subsection{Image Current} %%%%%%%%%%%%
  \label{Basic.2}
   %{\S\bf General shape}\\ 
\p{The lack of significant photospheric magnetic flux evolution} during the initiation of a CME can be modeled with the introduction of image currents below the photosphere.   The straight channel case of \citet{Kuperus74} can be generalized to any channel shape as illustrated in Figure~\ref{Fig_schema}(a).
Let the current vector be $(I_x,I_y,I_z)$ at a generic point $(x,y,z)$ of the corona, then the introduction of the image current $(-I_x,-I_y,I_z)$ at the image position $(x,y,-z)$ implies that the vertical component of the magnetic field at the photosphere, $B_z(x,y,0)$, is unaffected by the presence of the current channel.

   %{\S\bf general theory of images}\\ 
The introduction of this image current is a particular case of the technique of images to impose a particular boundary condition in the setting of electrodynamic problems \citep[][chap.~2 and 5]{Jackson75}.  Physically, the coronal current path is closing in a complex set of horizontal photospheric currents, which create the same coronal magnetic field than the image current.  
%(unicity of the solution). 

\subsection{Magnetic flux} %%%%%%%%%%%%
  \label{Basic.3}

   %{\S\bf Why flux ?}\\ 
We suppose that photospheric evolution and magnetic reconnection are negligible during the instability phase, then the total coronal magnetic flux passing below the current channel is conserved.  This flux is the sum of the flux of $\Bp$ and of $\BI$.   By the symmetric construction of the image current, the coronal flux of $\BI$ is half the flux enclosed by the full current channel (coronal and image current), so it is equal to $L I /2$,
where $L$ is the \p{external} inductance of the full current channel \citep[e.g.][]{Jackson75}.   

   %{\S\bf Circular case}\\
The total inductance of a circular channel of main radius $h$ and of small radius $a$ (Figure~\ref{Fig_schema}(b)) is    
  \begin{equation} \label{L_c}
  \Lc = \mu_{0} h \left(\ln \left( \frac{8h}{a}\right) -2 + \frac{\li}{2} \right) \,,
  \end{equation}
where $\mu_{0}$ is the magnetic permeability, and $\li$ is the normalized internal inductance, per unit length, of the current channel \citep{Grover46}.  $\li$ take the value of $0$, $0.5$ or $1$ for a current concentrated at the border of the torus, uniformly distributed within the cross section, or with a linear force-free field equilibrium \citep[][and references therein]{Lin98}.  Equation~(\ref{L_c}) is simple, but still a good approximation of more complete expressions.  For example, it is close to the series expansion tabulated by \citet{Malmberg65} in the range $0.1< a/h <1$ for $\li=0$.    
Moreover, the expression with elliptic integrals given by \citet[][p. 193]{Ramo94} is also close to Equation~(\ref{L_c}) with $\li=0$ in the range $0.15 \leq a/h <1$, while for lower $a/h$ values it is closer to the case $\li=0.5$.   Finally, for the coronal magnetic flux passing below the current channel, the external inductance is required, so the flux is $\Lc I /2$ with $\li=0$.

   %{\S\bf straight case}\\
With the same notations (Figure~\ref{Fig_schema}(c)), the total inductance of a straight channel and its image, for a length $ \Delta y$ along the channel, is   
  \begin{equation} \label{L_s}
  \Ls = \frac{\mu_{0} \Delta y}{\pi} \left(\ln \left( \frac{2h}{a}\right) 
      + \frac{\li}{2} \right) \,,
%  \Ls = \frac{\mu_{0} \Delta y}{\pi} \left(\ln (2 h/a) + \li /2 \right) \,,
  \end{equation}
where $\li$ has the same value than in the above circular case.
The main difference with the circular case, is that $\Ls$ is only weakly dependent
of $h$ so that $\Ls$ is almost constant during a global perturbation of the channel (modifying the height $h$).  As above, the coronal magnetic flux passing below the current channel is $\Ls I /2$ with $\li=0$.

\subsection{Magnetic Self Force} %%%%%%%%%%%%
  \label{Basic.4}

   %{\S\bf Origin of the force}\\ 
From the Biot and Savart law, a current channel generates a magnetic field at any point of the space, in particular in the current channel.  The component of this field orthogonal to the current channel induces a Laplace force.   The direct calculation of this force from the Biot and Savart law is very cumbersome, even for a simple torus geometry.
In practice, this force is computed by equaling the work of this force to the change of the magnetic energy ($L I^2/2$) during an elementary displacement, preserving the magnetic flux encircled by the current channel, $LI$, so that there is no inductive effect \citep{Shafranov66,Garren94}.  For a circular current channel, this implies an outward radial force $f_{\rm c}$, per unit length along the channel, given by:
  \begin{equation} \label{f_c}
  f_{\rm c} = \frac{I^2}{2} \frac{\partial \Lc}{\partial h}
            = I^2 \frac{\mu_{0}}{4 \pi h} \left(\ln \left( \frac{8h}{a} \right) -1 
            + \frac{\li}{2} \right) \,.
  \end{equation}
The force $f_{\rm c}$ is the Laplace force between the toroidal (axial) current and the poloidal magnetic field.  There is also the Laplace force between the poloidal current and the toroidal magnetic field inside the twisted flux tube.  Taking into account the internal equilibrium, in the limit $a/h<<1$, this second force only gives a small correction to the previous force, as the total force is obtained from Equation~(\ref{f_c}) by replacing $\li$ by $\lf =\li-\sh $ with $\sh \approx 1$ \citep{Shafranov66}.  Then, we write the total force as $\rep_{\rm c} I^2$ with    
  \begin{equation} \label{rep_c}
  \rep_{\rm c} = \frac{\mu_{0}}{4 \pi h} \left( \ln \left( \frac{8h}{a}\right) -1 
               + \frac{\lf}{2} \right)
  \end{equation}

The local outward force $\rep_{\rm c}I^2$, called hoop force, has its origin in the magnetic field created by each of the elementary part of the circular ring, with the largest contribution coming from the closest currents.  Indeed, it has a logarithm divergence as the small radius, $a$, \p{becomes} smaller. For $a<<h$ the logarithm term slightly dominates in Equation~(\ref{rep_c}).  If the current channel is fully in the corona \citep[e.g.][]{Lin98}, Equation~(\ref{rep_c}) includes only the self-force of the coronal current channel, whereas if the current channel is only half in the corona \citep[e.g.][Figure~\ref{Fig_schema}(b)]{Lin02} Equation~(\ref{rep_c}) also includes the force from the image current.  Finally, \citet{Garren94} have generalized these results to current channels with arbitrary shapes. 

   %{\S\bf straight case}\\ 
For a straight current channel, located at the height $h$ above the photosphere,
the magnetic force can be computed as above (Equation~(\ref{f_c})) with \p{$\Lc$ replaced by} $\Ls$ or the magnetic field of the image current can be computed first by the Biot and Savart law. The
repulsion function $\rep_{\rm s}$ is   
  \begin{equation} \label{rep_s}
  \rep_{\rm s} = \frac{\mu_{0}}{4 \pi h} \,.
  \end{equation}
$\rep_{\rm s}$ has indeed a similar form as $\rep_{\rm c}$ (Equation~(\ref{rep_c})) since the logarithm term provides only a weak dependance on $h/a$.

%%%%%%%%%%%%%%%%%%%%%%%%%%%%%%%%%%%%%%%%%%%%%%%%%%%%%%%%%%%%%%%%%%%%%%%%%%%
\section{Loss of Equilibrium and Instability} 
  \label{Loss}
%\section{Examples depending of one spatial coordinate}

\subsection{Magnetic Field Evolution} %%%%%%%%%%%%
  \label{Loss.1}

   %{\S\bf Typical AR evolution}\\ 
The MHD evolution of the coronal magnetic field of an active region, outside flare times, has typically three time-scales, as follows. 
  The shortest time-scale, $\tau_{A}$, called the Alfv\'en time, is given by the typical time that Alfv\'en waves
require to cross the coronal field configuration. Outside of the eruption period, the magnetic configuration is not significantly changing on such time scale and Alfv\'en waves transport the magnetic stresses (shear, twist). 
  On an intermediate time-scale, $\tau_{B}$, the coronal magnetic stress is significantly changing, for example because of sub-photospheric torsional Alfv\'en waves are bringing twist to the coronal field, or because of reconnection between sheared magnetic loops is forced by converging flows at the PIL. On this intermediate time-scale, coronal currents significantly change, but not the global photospheric distribution of the magnetic field component normal to the photosphere (i.e. the ``normal magnetogram''). 
  On the longest time-scale, $\tau_{C}$, both the coronal currents and the ``normal magnetogram'' are evolving.  
 
   %{\S\bf Typical AR time-scales}\\ 
 For a mature AR, with a magnetic flux of $\approx 10^{22}$~Mx, the above time-scales are typically of the order of $\tau_{A} \approx $ few minutes, $\tau_{B} \approx $ few days, and a $\tau_{C} \approx $ week to months.  They are well separated,
a useful property for an analytical study since it allows to isolate the main physics involved for each time-scale.  Thus, we can use different approximations to analyze the magnetic field evolution on these time-scales, as follows.  
% Apart the dissipation in small scale events (giving in particular a coronal heating), 
The global stability of the equilibrium can be tested with ideal MHD on the shortest time-scale $\tau_{A}$.   On the intermediate time-scale $\tau_{B}$, the build up of coronal current can be studied within a given potential $\Bp$ (computed uniquely from the ``normal magnetogram'', using appropriate boundary conditions on the sides of the domain).  Finally the evolution on the longest time-scale $\tau_{C}$ is mostly studied with MHD simulations or with analytical studies imposing an ideal-MHD evolution.

\subsection{Loss of Equilibrium with a Circular Current Channel}
  \label{Loss.2}
   %{\S\bf summary of the model}\\ 
    First, we revisit the model proposed by \citet{Titov99}, keeping it the simplest as possible.  The potential field $\Bp$ is created by two magnetic sources, of equal flux $\phi$ but of opposite sign, located at $x = \pm D, y=z=0$ and by a line current located at $y=z=0$.  The magnetic field of this line current adds a contribution to the coronal potential field, but it does not modify the equilibrium, nor the equilibrium perturbation studied below, so this field is not present in the following equations (the aim of this magnetic field is only to have a finite twist in the coronal field).  A torus of electric current, centered at $x=y=z=0$, is introduced in the plane y-z (Figure~\ref{Fig_schema}(b)).  Since the photospheric boundary is set at $z=0$, the half part of the torus at $z<0$ represents the image current. 
    
   %{\S\bf equilibrium}\\
The magnetic configuration considered is axisymmetric (around the x-axis), then the magnetic force on the current channel is radial in the y-z plane with the same value along the current channel:   
  \begin{equation} \label{force}
  f =  \rep_{\rm c}  \, I^2 + \Bpx  \, I  \,,
  \end{equation}
where $\rep_{\rm c}$ is given by Equation~(\ref{rep_c}).   
The x-component of $\Bp$ at the current location is:
  \begin{equation} \label{bx_c}
  \Bpx = - 4 \phi D (h^2+D^2)^{-3/2}   \,.
  \end{equation} 
The equilibrium current, $I_{\rm eq.}$ is given by $f =0$ in Equation~(\ref{force}):
  \begin{equation} \label{Ieq_c}
  I_{\rm eq.}(h) = \frac{16 \pi}{\mu_{0}} 
                \frac{\phi D h (h^2+D^2)^{-3/2}}{\ln(8 h/a) -1 + \lf/2}    \,. 
  \end{equation}
  %$I_{\rm eq.}(h)$ define the equilibrium curve and it has typically one maximum.
Starting from a nearly potential configuration (small $I$ value), if the current value could be progressively increased in function of time as in a classical electric circuit, then a loss of equilibrium would
occur when $I$ would reach the maximum value of $I_{\rm eq.}(h)$.   The circular current channel behaves in the same way as a straight current channel and its image current (Figure~\ref{Fig_schema}(c) and Section~\ref{Loss.6}) as proposed by \citet{vanTend78}.    However, with an MHD evolution, the magnitude of the coronal current is rather determined by the magnetic field evolution, so its time evolution cannot be imposed a priori. Indeed, during a typical MHD evolution with imposed photospheric velocities, the coronal current magnitude first grows then later decreases \citep[e.g.][]{Aly85,Klimchuk89,Aulanier05}.   
  
   %{\S\bf evolution}\\
In previous studies, the magnetic field evolution is typically assumed to be ideal \citep[e.g.][and references therein]{Isenberg07}. With no magnetic flux emerging or canceling between the photospheric sources (located at $x=\pm D$), this implies the conservation of the magnetic flux, $F$, through the area, S, defined between $z=0$ and the bottom of the current channel. $F$ is given by
  \begin{eqnarray}
  F &=& \frac{\Le I}{2} + {\int \!\!\!\! \int}_{\!\!\rm S} \Bpx \,\rmd y \,\rmd z  \label{flux}  \\
    &=& \frac{\Le I}{2} 
       - 4\pi \phi \left(1 - \frac{D}{\sqrt{(h-a)^2+D^2}} \right)     \,, \label{flux_c} 
  \end{eqnarray}    
with $\Le=\mu_{0} h (\ln 8 h/a -2)$ is the external inductance of the current channel (see Equation~(\ref{L_c})).
The conservation of $F$, together with the time evolution of one parameter of the model (e.g. $\phi$ or $D$) provides an evolution constraint with the generic form $I_{\rm evol.}(h)$.  Its intersection with the equilibrium curve, $I_{\rm eq.}(h)$, determines the evolution of $I$ in function of the evolving parameter (e.g. $\phi$ or $D$).

%%%%%%%%%%%%%%%%%%%%%%%%%%%%%%%%%%%%%%%%%%%%%%%%%%%%%%%%%%%%%%%%%%%%%%%%%%
\begin{figure}[t!]    
\IfFileExists{submit.txt}{
  \centerline{\includegraphics[width=0.5\textwidth, clip=]{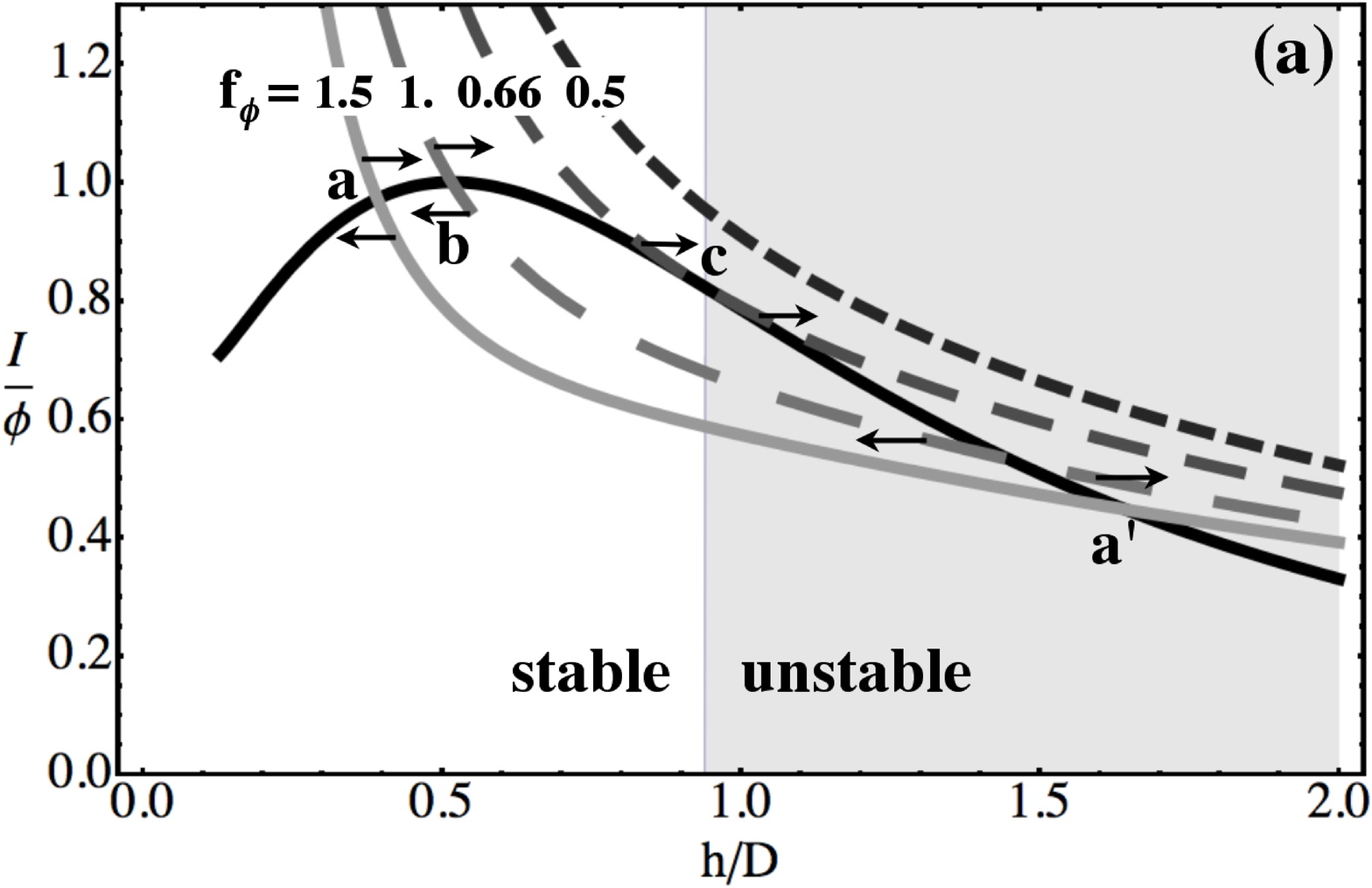}
              \includegraphics[width=0.5\textwidth, clip=]{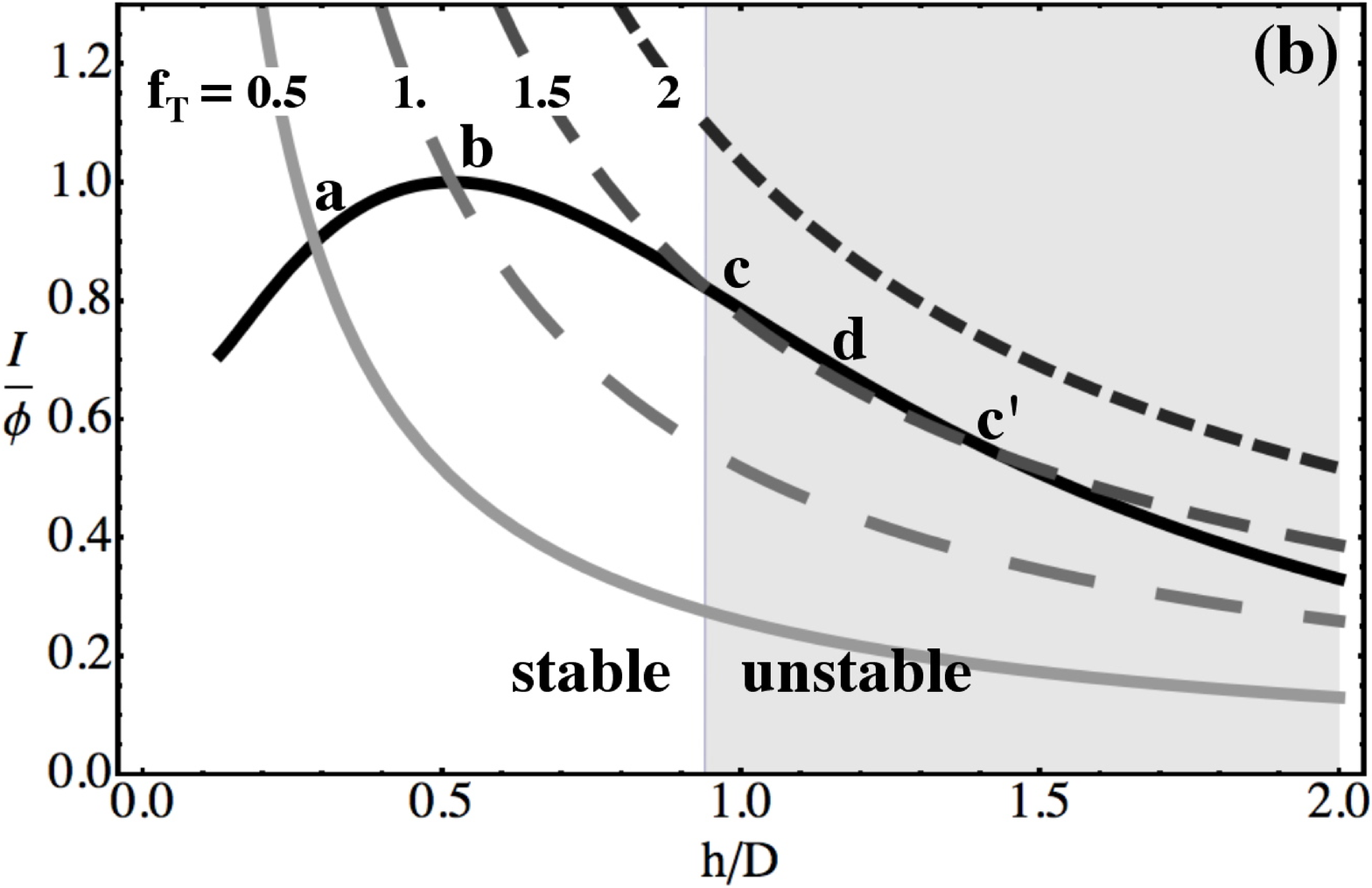}}
  \centerline{\includegraphics[width=0.5\textwidth, clip=]{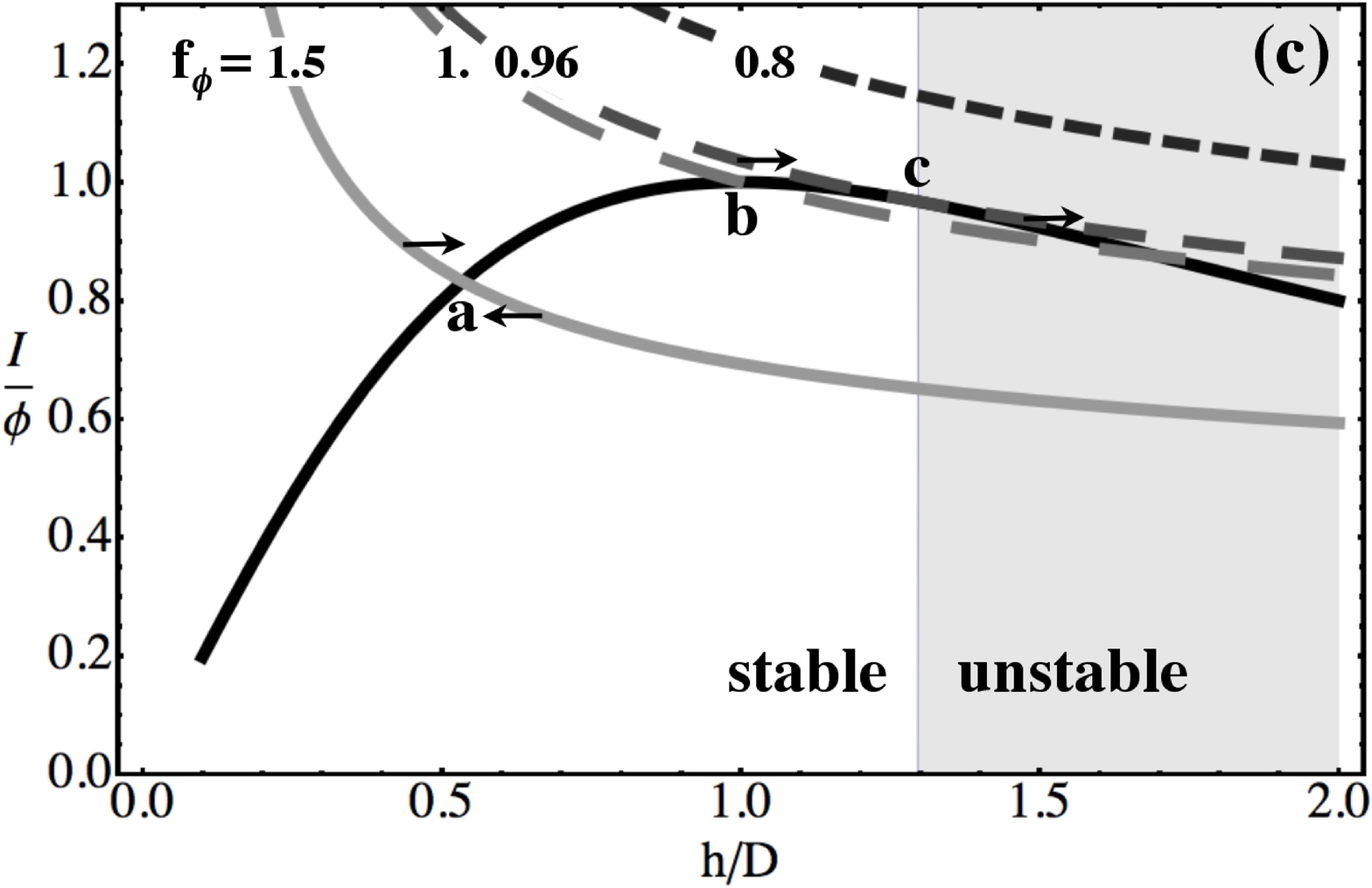}}
}{
  \centerline{\includegraphics[width=0.45\textwidth, clip=]{f2_lost_of_equil_a}}
  \centerline{\includegraphics[width=0.45\textwidth, clip=]{f2_lost_of_equil_b}}
  \centerline{\includegraphics[width=0.45\textwidth, clip=]{f2_lost_of_equil_c}}
} 
\caption{The coronal electric current, $I$, divided by the photospheric magnetic flux $\phi$, in function of the coronal height $h$, normalized to half the $x$ distance, $D$, of the photospheric magnetic polarities. The equilibrium curve, $I_{\rm eq.}/\phi(h)$, is in black, while the evolution constraint is shown in grey/dashed style for four values of the evolution factor $f$. Both $f$ and $I_{\rm eq.}/\phi$ are normalized to $1$ at the maximum of $I_{\rm eq.}/\phi$. 
  {\bf (a,b)} Case of a circular current channel (Figure~\ref{Fig_schema}(b)), 
  {\bf (c)} case of a straight current channel (Figure~\ref{Fig_schema}(c)). 
  {\bf (a,c)} The photospheric flux, $\phi$, is evolved with the factor $f=f_{\phi}$. 
  {\bf (b)} The magnetic twist of the flux tube is evolved with the factor $f=f_{\rm T}$.  Arrows represent the direction of the force when the system is perturbed away from the equilibrium. 
}
 \label{Fig_non_equil}
\end{figure}

   %{\S\bf equilibrium + evolution}\\
With a circular current channel, \citet{Lin02} has shown that evolving $D$, with $F$ preserved, does not lead to a loss of equilibrium, but rather to a self similar evolution of the configuration (which is just a rescale in size).  However, they also show that a decrease of $\phi$ does lead to a loss of equilibrium.   Figure~\ref{Fig_non_equil}(a) is a graphical representation of this result.  As in previous studies, the maximum of $I_{\rm eq.} (h)$ defines the reference state with fluxes $\phi_0$ and $F_0$.  The evolution is parametrized by $f_{\phi}$ defined by 
$\phi= f_{\phi}\, \phi_0$, where $f_{\phi}$ is a decreasing function of time as observed after the emerging phase of an active region \citep[e.g.][]{vanDriel03} or before filament eruption \citep[e.g.][]{Schmieder08}.  In order to have an equilibrium curve not evolving with $\phi$ (to simplify the graphic), we draw $I_{\rm eq.}/\phi$ (normalized to its maximum value), as well as $I_{\rm evol.}/\phi$. 

   %{\S\bf stable equilibrium}\\
Starting the evolution with $f_{\phi}>1$, the constraint $F=F_0$ has an intersection with the equilibrium curve before its maximum (e.g. at the \pta\ for $f_{\phi}=1.5$ in Figure~\ref{Fig_non_equil}(a)).  This equilibrium is stable since the perturbed equilibrium, with the constraint $F=F_0$, has a restoring force as shown with the arrows (see Section~\ref{Stab.1} for the analysis of the perturbed equilibrium).  

   %{\S\bf loss of equilibrium}\\
As $f_{\phi}$ decreases (due e.g. to magnetic flux cancelation with some photospheric flux brought from $|x|>D$), the equilibrium height $h$ increases.  Contrary to what is obtained if the evolution of $I$ is prescribed as in a circuit model, the conservation of magnetic flux in ideal MHD leads to the equilibrium to be still stable after the maximum of $I_{\rm eq.}/\phi$ (\ptb), up to the \ptc. 

 For larger $h$ values, e.g. at \ptap , the equilibrium is unstable to an ideal perturbation (i.e with $F$ preserved).  However, this unstable region is not reachable during the pre-eruptive evolution: the quasi-static evolution ends at \ptc\ with both an equilibrium becoming unstable (see arrows in Figure~\ref{Fig_non_equil}(a)) and without neighbor equilibrium, as $f_{\phi}$ is slightly more increased, so there is a loss of equilibrium.  The evolution sequence with decreasing $f_{\phi}$ ends with a fast evolution driven by an outward force, so an eruption.  It could be confined if a stable equilibrium would be present at greater height, e.g. due to the formation of a long current sheet below the flux rope \citep{Forbes91}, or if a too strong magnetic tension of the overlying field would be present \citep[as for the kink-unstable eruption modeled by][]{torok05}.

\subsection{Other Possible Evolutions on Intermediate Time Scales}
  \label{Loss.3}

   %{\S\bf More general evolution}\\
The analysis of the previous section supposes an ideal-MHD evolution to progressively shift the equilibrium to a point of the equilibrium curve where no neighbor equilibrium exists (with the magnetic flux conservation).  This is only one plausible scenario for solar eruptions (see Section~\ref{Introduction}).  Indeed, on the intermediate time-scale $\tau_{B}$, magnetic reconnection can play a crucial role in transforming the coronal field.  For example, this is the case with a progressive diffusion of photospheric magnetic polarities leading to reconnection at the PIL and the transformation of sheared to twisted field lines,  contributing to build up the twisted flux tube and the associated current channel \citep[as in the numerical simulations of][]{Amari03b,Fan04,Mackay06,Aulanier10}.  A progressive transformation of the coronal magnetic configuration is also expected due to reconnection of the current layers formed at separatrices and Quasi-Separatrix Layers \citep[QSLs, e.g.][]{Titov02} during the intermediate time-scale $\tau_{B}$.   We conclude that typically it is not obvious to justify an ideal-MHD evolution during the long pre-eruptive build-phase before an eruption, on time scales much longer than $\tau_{A}$, and in particular using the evolution constraint of a preserved flux below the flux rope (e.g. Equation~(\ref{flux_c})). 

   %{\S\bf More general evolution}\\
\p{More generally, for a given observed configuration or in a 3D MHD simulation, it is difficult to
determine the main evolutionary constraint, even with detailed analysis \citep[e.g.][]{Aulanier10}. This is not due to a lack of available data; rather, the difficulty is
due to the complexities inherent in the 3D evolution of magnetic fields.}
This is illustrated by the formidable complexity of studying the loss of equilibrium even in a simplified configuration \citep[e.g. a small bipole emerging in a bipolar field,][]{Lin01}. 
 
\subsection{Instability with a Circular Current Channel}
  \label{Loss.4}

   %{\S\bf More general evolution}\\
Let us illustrate another possible evolution than the ideal-MHD evolution on the time scale $\tau_{B}$.  This following case is selected mainly because of its simplicity to illustrate other possible evolutions.  The current channel is associated with a twisted flux tube which has a finite twist due to the presence an axial magnetic field. %of line current (located at $y=z=0$). 
The average coronal twist, $T$, is approximately related to the current $I$ and the toroidal flux, $\phi_{\rm t}$, in the flux rope by \citep[see Equation(9) in][]{Titov99}: 
  \begin{equation} \label{twist}
  T = \mu_{0} I h ~/~ (2 \phi_{\rm t})     \,.
  \end{equation}
Let us suppose that the flux rope twist $T$ is increasing, e.g. due to torsional Alfv\'en waves
coming from the convective zone, or as a consequence of reconnecting sheared loops (increasing the flux rope flux). Equation~(\ref{twist}) provides a new evolution constraint, replacing the conservation of $F$ used in Section~\ref{Loss.2}.  As previously, the maximum of $I_{\rm eq.} (h)$ defines the reference state with the flux $\phi_0$ and the twist $T_0$.  Here, we simply suppose that the photospheric field sources are not evolving so $\phi=\phi_0$, but we still plot $I_{\rm eq.}/\phi$ for coherence with previous case.  The evolution is parametrized by $f_{\rm T}$ defined by $T= f_{\rm T} \, T_0$. 

   %{\S\bf Describe evolution}\\
Starting the evolution with $f_{\rm T}<1$, the evolution constraint, Equation~(\ref{twist}), has an intersection with the equilibrium curve before its maximum (e.g. at the \pta\ for $f_{\rm T}=0.5$ in Figure~\ref{Fig_non_equil}(b)).  As in Section~\ref{Loss.2}, this equilibrium is stable.  As $f_{\rm T}$ increases, the equilibrium height $h$ increases, reaching \ptb\ then \ptc.  If we only consider the evolution constraint, the evolution would reach \ptd, where no neighbor equilibrium is present when $f_{\rm T}$ is further increased. As in the previous case (in Figure~\ref{Fig_non_equil}(a)) a loss of equilibrium is present, but at larger height.  

However, the true physical evolution of the system is ending earlier, at \ptc, where the system is unstable on the short time scale $\tau_{A}$ with an ideal-MHD perturbation (preservation of the flux $F$ in Equation~(\ref{flux_c}), while the evolution of $T$ is negligible on the short time scale $\tau_{A}$).

\subsection{Loss of Equilibrium or Instability?}
 \label{Loss.5}

   %{\S\bf Comparison of the evolutions}\\
In the two previous evolutions (Figures~\ref{Fig_non_equil}(a) and (b)) the eruption occurs at the same location along the equilibrium curve, at the \ptc.  However, they appear to have a different theoretical origin.  In the first case (Figure~\ref{Fig_non_equil}(a)) the progressive decrease of $f_{\phi}$ brings the system to \ptc\ where no neighbor equilibrium exits with a further decrease of $f_{\phi}$.  In the second case (Figure~\ref{Fig_non_equil}(b)) the increase of $f_{\rm T}$ brings
also the system to \ptc, but a neighbor equilibrium is present with a further increase of $f_{\rm T}$ (up to \ptd).  Simply, the equilibrium after \ptc\ is unstable with an ideal perturbation ($F$ preserved), and as in the first case, an upward eruption is present after the system reaches \ptc.  Indeed, on the time scale $\tau_{A}$, the evolution of the two cases will be the same after they reach \ptc, since the same unstable force (Equation~(\ref{force})) is acting with the constraint of an ideal-MHD evolution (so preserving $F$ in Equation~(\ref{flux_c})).  Then, we argue that there is no point to discuss \p{whether} there is a loss of equilibrium or a transition from a stable to an unstable equilibrium.

   %{\S\bf Generalisation}\\
More generally, for an observed coronal field evolution, and even for an MHD simulation (where all physical quantities are available in the volume), we claim that, in most cases, it will be at least difficult, if not impossible, to define precisely the evolution constraint as in the previous two cases.  Indeed, with a slow enough driving, the system follows the equilibrium curve, and the extensions away from this curve, along the evolution constraint, are purely theoretical considerations which are available only if an analytical analysis is achievable.  In a magnetic configuration which includes some separatrices or QSLs, so with some reconnection, we claim that an evolution constraint cannot generically be constructed. 
However, a generic approach is to test the ideal-MHD stability all along the evolution (within the limit of numerical dissipation for MHD simulations). 

   %{\S\bf Cannot start from the unstable equilibrium}\\
Because of the separation of the time scale of the coronal evolution ($\tau_{A}$)
from the longer times scales of the photospheric evolution ($\tau_{B}$, $\tau_{C}$, see Section~\ref{Loss.1}), the magnetic configuration cannot reach the equilibrium
branch at an altitude $h$ larger than that of the point ``c'' by a slow quasistatic evolution, which does correspond to observed pre-eruptive evolutions.  Reaching a region of the equilibrium curve beyond ``c'' may be dynamically possible, but only for
fast (e.g. Alfv\'enic) evolutions of magnetic configurations being out of equilibrium. This is, however, not observed in the Sun's atmosphere, even during flux emergence. Nevertheless, considering an analytical model of the equilibrium, one can always start a numerical simulation from any point along the equilibrium curve, stable or unstable
\citep[e.g.][]{Torok07}. The unstable branch is unaccessible for a coronal field, except the vicinity of point ``c'' which can be reached with a small but finite velocity due to the slow evolution present in the pre-eruptive stage. Indeed, including such small velocity implies an evolution curve, $h(t)$, which is in better agreement to observed $h(t)$ in prominence eruptions, than letting the instability grows from an initial very small perturbation \citep{Schrijver08}.

\subsection{Loss of Equilibrium with a Straight Current Channel}
  \label{Loss.6}

   %{\S\bf More general cases}\\
The evolution summarized Figures~\ref{Fig_non_equil}(a) and (b) is expected to be generic of magnetic configurations having at least one current channel which is not fully neutralized since the hoop force is generically present with a curved current channel \citep{Garren94}.  Indeed, this has been shown in more complex configurations, even by including a complete photospheric line tying, i.e. not only fixing the ``normal magnetogram'' (with the inclusion of image current, see Section~\ref{Basic.2}), but also fixing the photospheric positions of the current channel \citep{Isenberg07}.  

   %{\S\bf Why studing another case}\\
  However, the above physical evolution is not limited to the presence of the hoop force.  It is indeed generic of the Lorentz force created by any current channel.  At a given position of the circuit, the magnetic force can be dominated either by the magnetic field created by the current located in the vicinity of this position (hoop force) or by the current at large distance (e.g. an image current).  Indeed, \citet{vanTend78} first proposed a catastrophe model of a straight current channel embedded in a potential field.  
%The image current (or its equivalent current distribution at $z=0$) creates a coronal magnetic field with a Lorentz force, which acts on the current channel, equal to $\rep_{\rm s}\, I^2$, with $\rep_{\rm s}$ defined by Equation~(\ref{rep_s}).

   %{\S\bf equilibrium}\\
As in Section~\ref{Loss.2}, a potential field $\Bp$ is introduced to achieve an equilibrium. 
As previously, we select a bipolar field $\Bp$ created by two magnetic sources of flux $\phi$ located at $x = \pm D, z=0$, but now invariant by translation in the $y$ direction.
The x-component of $\Bp$ at the current location is:
  \begin{equation} \label{bx_s}
  \Bpx = - 2 \phi D (\pi (h^2+D^2))^{-1}     \,.
  \end{equation}
The equilibrium current, $I_{\rm eq.}$ is given by $f =0$ in Equation~(\ref{force}):
  \begin{equation} \label{Ieq_s}
  I_{\rm eq.}(h) = \frac{8 \pi}{\mu_{0}} 
                \frac{\phi D h}{h^2+D^2}      \,.
  \end{equation}
The equilibrium curve is closely similar to the one found for a circular current channel
(compare panels a and c of Figure~\ref{Fig_non_equil}).  The main difference is that the maximum of $I_{\rm eq.}$ is shifted to a larger height.  This is mainly
due to the different dependance with $h$ of $\Bpx$ for a 2D and 3D bipole. 
Much closer equilibrium curves are obtained when the 2D bipole is replaced by a 2D quadrupole
(giving a potential-field dependance similar to Equation~(\ref{bx_c})).

   %{\S\bf evolution}\\
The conservation of the magnetic flux below the current channel, per unit length along its axis (Equation~(\ref{flux})), for the  straight channel case is:
  \begin{equation}
  F = \frac{\mu_{0} I}{2 \pi} \ln \frac{2 h}{a}  
       - \frac{2 \phi}{\pi} \tan^{-1}\frac{h-a}{D}  \,. \label{flux_s} 
  \end{equation}    
A major difference with the circular case is that the constraint of $F$ conservation implies that $I$ has a much weaker dependence on $h$ (Figure~\ref{Fig_non_equil}(c)).  It implies that the ideal-MHD instability occurs just after the maximum of $I_{\rm eq.}/\phi$ (where $f_{\phi}=1$), when the photospheric polarities have weaken by only 4\% ($f_{\phi}=0.96$).   Indeed, the ideal-MHD evolution leads to a non equilibrium at a location nearby to the one found by \citet{vanTend78} with an evolution simply driven by an increase of the current $I$.   
 
   %{\S\bf non equilibrium/instability}\\
As for the circular current channel, a loss of equilibrium at \ptc\ is only present if the magnetic evolution is fully ideal. More generally, whatever is the driver of the evolution on the time scale $\tau_{B}$, the system is becoming unstable as it reaches \ptc, and it is ideally driven away from the equilibrium curve by the same force in the short time scale $\tau_{A}$. 

\subsection{Comparison to Previous Studies}
  \label{Loss.7}

   %{\S\bf  van Tend loss of equilibrium / torus instability}\\
 Based on our analysis above, we can now answer to the following question: is there a major difference in the following approaches: ``loss of equilibrium'' studies with straight or circular current channel \citep[e.g.][]{vanTend78, Lin02}, and the ``torus instability'' \citep{Kliem06} previously studied in the tokamak laboratory experiment \citep[e.g.][]{Bateman78}?

   %{\S\bf  Similar repulsion function }\\
First, the straight or circular current channels have very similar repulsion forces, $r(h) I^{2}$,
implying the same kind of equilibrium curve.  Indeed, for a circular current channel, $r_{\rm c}(h)$ has a contribution from both the coronal part ($z>0$) and from the image current ($z<0$). Simply, both contributions can be combined in a single term, Equation~(\ref{rep_c}), masking the contribution of the image current.   More generally, \citet{Garren94} have derived a general
expression for $r(h)$ for arbitrary current shapes. $r_{\rm c}$ and $r_{\rm s}$ (Equations~(\ref{rep_c}) and (\ref{rep_s})), are simply two limits of the same $r(h)$, for circular and straight current channels.  

   %{\S\bf  loss of equilibrium / torus instability}\\
 Next, \citet{Kliem06} studied the ``torus instability'' of a circular current channel
imposing a constant current $I$ or a constant flux $F$.  The first \p{case is directly comparable to the work of \citet{vanTend78} and the second case to the work of \citet{Lin02}.}  The only significant difference is that \citet{Kliem06} study the stability of the equilibrium curve (Figure~\ref{Fig_non_equil}(a)), but do not follow the evolution of the magnetic configuration on intermediate time scale $\tau_{B}$ (so they cannot detect the presence or not of a loss of equilibrium).  

   %{\S\bf  Conclusion}\\
The main difference between the loss of equilibrium and the stability analysis is that, for the first case, a precise way how the system can evolve to instability is proposed, while, in the second case, one only tests the stability of a given equilibrium (which could be physically inaccessible).  However, with both type of analyses, an ideal instability is present at the same location of the equilibrium curve if the same equilibrium is analyzed.  This last condition is not trivial if one allows the formation of current sheets during the long-term evolution (on the time scale $\tau_{B}$).

%%%%%%%%%%%%%%%%%%%%%%%%%%%%%%%%%%%%%%%%%%%%%%%%%%%%%%%%%%%%%%%%%%%%%%%%%%%
\section{Equilibrium Stability} 
  \label{Stab}

  We study below the stability of the magnetic configuration around an equilibrium position.  The main \p{assumption} is that the magnetic force balance is the same at any position along the current channel, so that the magnetic force can written as in Equation~(\ref{force}).  This includes both circular and straight current channels in the same formalism. 
The stability of the configuration is first derived in this general framework in Section~\ref{Stab.1}, with the constraint of magnetic flux conservation derived in Section~\ref{Stab.2}.
These results are used for the particular cases studied in previous section.  Finally a parametric study of the stability is presented.   

\subsection{Force Perturbation} %%%%%%%%%%%%
   \label{Stab.1}
The magnetic force, $f(a,h,I)$, on the current channel is \p{described} by Equation~(\ref{force}). A perturbation $(\rmd a, \rmd h, \rmd I)$ around the equilibrium, $f(a,h,I)=0$, creates the force
$\rmd f$ as 
  \begin{equation}  \label{df-pert}
  \rmd f = \frac{\partial f}{\partial a} \rmd a 
         + \frac{\partial f}{\partial h} \rmd h
         + \frac{\partial f}{\partial I} \rmd I      \,.
  \end{equation} 
    
During a perturbation the radius $a$ is evolving as given by the internal force balance.  With a linear force-free field inside the flux rope and ideal MHD, \citet{Lin98} found that $a$ evolves as $1/I$.  Here we include a more general variation, supposing $a(I)$, and we introduce the index of variation of $a$ with $I$
  \begin{equation}  \label{na}
  \na \equiv  - \frac{\partial \ln a }{\partial \ln |I|}     \,,  
  \end{equation}   
so $\na =1$ for the internal evolution included in \citet{Lin98}.  Then, Equation~(\ref{df-pert}) is rewritten as 
  \begin{equation}  \label{df-pert1}
  \rmd f = \left( \frac{\partial f}{\partial h}  
         +   \left(-\na \frac{a}{I} \frac{\partial f}{\partial a} 
                                  + \frac{\partial f}{\partial I} \right)
         \left. \frac{\rmd I}{\rmd h} \right| _{\rm pert.}  \right) \rmd h   \,,
  \end{equation} 
where $\rmd I/\rmd h | _{\rm pert.}$ express how the current intensity is modified during the perturbation. 

If the perturbation is realized along the equilibrium curve, the force $\rmd f$ has the same expression except that $\rmd I/\rmd h | _{\rm eq.}$, computed along the equilibrium curve, replaces $\rmd I/\rmd h | _{\rm pert.}$ in 
Equation~(\ref{df-pert1}).  Also in this case, $\rmd f =0$ so  
  \begin{equation}  \label{df-eq}
      \frac{\partial f}{\partial h}  
         +   \left(-\na \frac{a}{I} \frac{\partial f}{\partial a} +\frac{\partial f}{\partial I}\right)
         \left. \frac{\rmd I}{\rmd h} \right| _{\rm eq.}  =0 \,,
  \end{equation}
where we suppose that the internal equilibrium has the same $\na$ index. 
Using the equilibrium condition, $f=\rep \, I^2 + \Bpx I =0$, and Equation~(\ref{df-eq}), %the perturbation of the magnetic force around the equilibrium (
Equation~(\ref{df-pert1}) is rewritten as      
  \begin{equation}  \label{df-pert2}
  \rmd f = r\,I \left( 1-\na \frac{\partial \ln r}{\partial \ln a} \right)
          \left( \left. \frac{\rmd I}{\rmd h} \right| _{\rm pert.} 
                -\left. \frac{\rmd I}{\rmd h} \right| _{\rm eq.}  \right) \rmd h     \,. 
  \end{equation}   
The equilibrium is unstable when $\rmd f$ and $\rmd h$ have the same sign. Then, an instability is present when the absolute value of the current in the perturbation decreases less rapidly with height than along the equilibrium curve.  This is illustrated in Figure~\ref{Fig_non_equil}.
 % (with $\phi$ constant during the perturbation). 

   %{\S\bf Link to previous Section and to Figure~\ref{Fig_non_equil}}\\
On the short time-scale $\tau_{A}$, the perturbation is described by ideal MHD, with a preservation of the flux distribution at the boundary $z=0$ (so $\phi=$constant in the examples of Section~\ref{Loss.2} and \ref{Loss.6}). Then, a perturbation corresponds to a small excursion away from the equilibrium curve with the constraint of magnetic flux conservation in Figures~\ref{Fig_non_equil}(a) and (c).  The tangent \ptc\ between the equilibrium curve and the constraint of flux conservation define both the limit of the stable region and the non equilibrium point (in an ideal evolution on the time-scale $\tau_{B}$). 

   %{\S\bf introduce decay index }\\
Using the equilibrium condition to \p{specify} $\rmd I/\rmd h|_{\rm eq.}$, the force perturbation, Equation~(\ref{df-pert1}), is rewritten as
  \begin{equation}  \label{df-pert3}
  \frac{\rmd f}{\rmd h} = \rep \, I \left( 1-\na \frac{\partial \ln r}{\partial \ln a} \right)
                \left. \frac{\rmd I}{\rmd h} \right| _{\rm pert.} 
                 + I \frac{\partial \Bpx}{\partial h} + I^2 \frac{\partial r}{\partial h}     \,.
  \end{equation}   
We define the decay index of the potential field as
  \begin{equation}  \label{nBp}
  \nBp \equiv - \frac{\partial \ln |\Bpx |}{\partial \ln h}    \,.   
  \end{equation}   
This decay index is introduced in Equation~(\ref{df-pert3}) by dividing $\rmd f$ by $\Bpx I$, or equivalently by $-\rep\, I^2$, and by multiplying by $h$.  Finally, the equilibrium is unstable (i.e. $\rmd f /\rmd h >0$) if 
  \begin{eqnarray}  
  \nBp \,>\, \nBpcrit &\equiv& \nr + \nI \nonumber \\
       &\equiv& - \frac{\partial \ln r}{\partial \ln h}
        - \left( 1-\na \frac{\partial \ln r}{\partial \ln a} \right)
          \left. \frac{\rmd \ln |I|}{\rmd \ln h} \right| _{\rm pert.}    \label{unstable}  % \,,
  \end{eqnarray}   
so if the potential field decreases fast enough with height.  The instability threshold depends on how fast the repulsion decreases with height (given by the decay index $\nr$) and on how much the current is allowed to decrease during the perturbation (given by the decay index $\nI$). 

\subsection{Constraint of Flux Conservation} %%%%%%%%%%%%
   \label{Stab.2}

   %{\S\bf flux perturbation}\\
The decay index $\nI$ is computed from an ideal MHD constraint, as follows. On the short time-scale $\tau_{A}$, the perturbation is described by ideal MHD, with the preservation of the coronal magnetic flux $F$ present below the flux rope.  With a small perturbation $\rmd h$, Equation~(\ref{flux}) implies
  \begin{equation} \label{dflux}
  \rmd F = \left( \frac{\rmd \Le}{\rmd h} I
           + \left. \Le \frac{\rmd I}{\rmd h}\right| _{\rm pert.}  
           + P~\Bpx           \right)\frac{\rmd h}{2}
         = 0      \,,  
  \end{equation}
where $P$ is the perimeter (or length) of the full current channel, including its image.

   %{\S\bf introduce the dependance of $\Le$}\\
The external inductance $\Le$ (computed with $\li=0$) depends generically on the spatial extension of the current channel  (described here only by $h$) and on the thickness of the current channel (its radius $a$). 
Then, the variation of the external inductance with height is
  \begin{eqnarray}
  \frac{\rmd \Le}{\rmd h} &=& \frac{\partial \Le}{\partial h} 
       + \frac{\partial \Le}{\partial a} \frac{\partial a}{\partial I}
            \left. \frac{\rmd I}{\rmd h}\right| _{\rm pert.} \nonumber \\
                          &=& \frac{\partial \Le}{\partial h}
       - \na \Le \frac{\partial \ln \Le}{\partial \ln a} 
            \left. \frac{\rmd \ln |I|}{\rmd h}\right| _{\rm pert.}     \,.   \label{dLe}
  \end{eqnarray}

   %{\S\bf Constraint of flux conservation on $I$ changes}\\
The conservation of the magnetic flux, Equation~(\ref{dflux}), together with Equation~(\ref{dLe}) imply the following decay index    
  \begin{equation} \label{n_I}
  \nI =  
     \frac{1-\na \frac{\partial \ln r}{\partial \ln a} }
         {1-\na \frac{\partial \ln \Le}{\partial \ln a} } 
     \left( \frac{\partial \ln \Le}{\partial \ln h} - \frac{P\,h\,\rep}{\Le} \right) \,.
  \end{equation}
The last factor in the previous equation is due to the conservation of the magnetic flux from the current $I$ minus the flux from the potential field passing below the current channel.   The fraction includes the effect of the channel expansion on the repulsion function \p{at the numerator and on the conservation of flux at the denominator.}  They are both larger than 1 for $\na>0$, and they are a growing function of $\na$. Then, the dependance of $r$ and $\Le$ on $a$  have opposite effects on the stability.

\subsection{Example of a Circular Current Channel} %%%%%%%%%%%%
   \label{Stab.3}
   %{\S\bf Contributions of Le to $I$ change with flux conservation}\\
  We apply in this subsection the above results to a circular current channel (Figure~\ref{Fig_schema}(b)). The external inductance $\Le$ is given by Equation~(\ref{L_c}), without the contribution of the internal inductance ($\li =0$).  The repulsion $r$ is given by Equation~(\ref{rep_c}) and $P=2\pi h$.  The contribution from the potential field to the decay index $\nI$ is (for $\na=0$, i.e. for $a$ constant during the evolution):  
  \begin{equation} \label{dlogI_Bp}
  - \frac{P\,h\,\rep}{\Le} = - \frac{1}{2} \left(1+ \frac{1+\lf /2}{\ln (8h/a) -2} \right)    \,.
  \end{equation}
   The contribution from the field created by the current channel to the decay index $\nI$ is (for $\na=0$)
     \begin{equation} \label{dlogI_channel}
  \frac{\partial \ln \Le}{\partial \ln h} = 1+ \frac{1}{\ln (8h/a) -2}     \,,
  \end{equation}
so exactly twice the negative contribution from the potential field for $\lf =0$. For $a \ll h$, i.e. a very thin current channel, the
denominator, $\ln (8h/a) -2$, is relatively large so that the above contributions are close to $-1/2$ and $+1$, respectively.  Including Equations~(\ref{dlogI_Bp},\ref{dlogI_channel}) in Equation~(\ref{n_I}), and using Equation~(\ref{L_c}), the decay index $\nI$ for the circular current channel is 
  \begin{equation}    \label{n_Ic}
  \nI = \frac{(\uf +\lf /2+\na)(\uf -\lf /2)}{2(\uf -1 + \na)(\uf + \lf /2)} \,,
  \end{equation}
with the notation
  \begin{equation}    \label{uf}
  \uf = \ln (8h/a) -1 \,.
  \end{equation}

   %{\S\bf Contributions of r to $I$ change with flux conservation}\\
Using Equation~(\ref{rep_c}), the decay index of the repulsion $r$ is:
  \begin{equation} \label{n_rc}
  n_r = - \frac{\partial \ln r }{\partial \ln h} = 1-\frac{1}{\uf +\lf /2}    \,.
  \end{equation}
From Equation~(\ref{unstable}), $n_r$ is the critical decay index of $\Bp$ for instability if the current $I$ would be preserved during the perturbation. $n_r$ is always lower than 1, especially for flux rope with large radius $a$. 

   %{\S\bf Critical nBp index}\\
Combining the results of Equations~(\ref{n_Ic},\ref{n_rc}), the instability condition for a circular current channel is  
  \begin{equation}    \label{unstable_c}
  \nBp > \frac{3}{2}  - \frac{(1+\lf /2)(\uf-2 +2\na +\lf /2)}
                             {2(\uf+\lf /2)(\uf-1+\na)} \,.
  \end{equation}
In the limit of a very thin current channel ($a \ll h$), the instability threshold is close to $3/2$, as found in tokamak studies \citep[e.g.][]{Bateman78}.   This corresponds to the ``torus instability'' for solar eruptions \citep{Kliem06}, but with a different correction term to $3/2$, as we have not supposed a self similar expansion of the current channel, but rather a dependance $a(I)$.   

%%%%%%%%%%%%%%%%%%%%%%%%%%%%%%%%%%%%%%%%%%%%%%%%%%%%%%%%%%%%%%%%%%%%%%%%%%
\begin{figure}[t!]    
\IfFileExists{submit.txt}{
  \centerline{\includegraphics[width=0.8\textwidth, clip=]{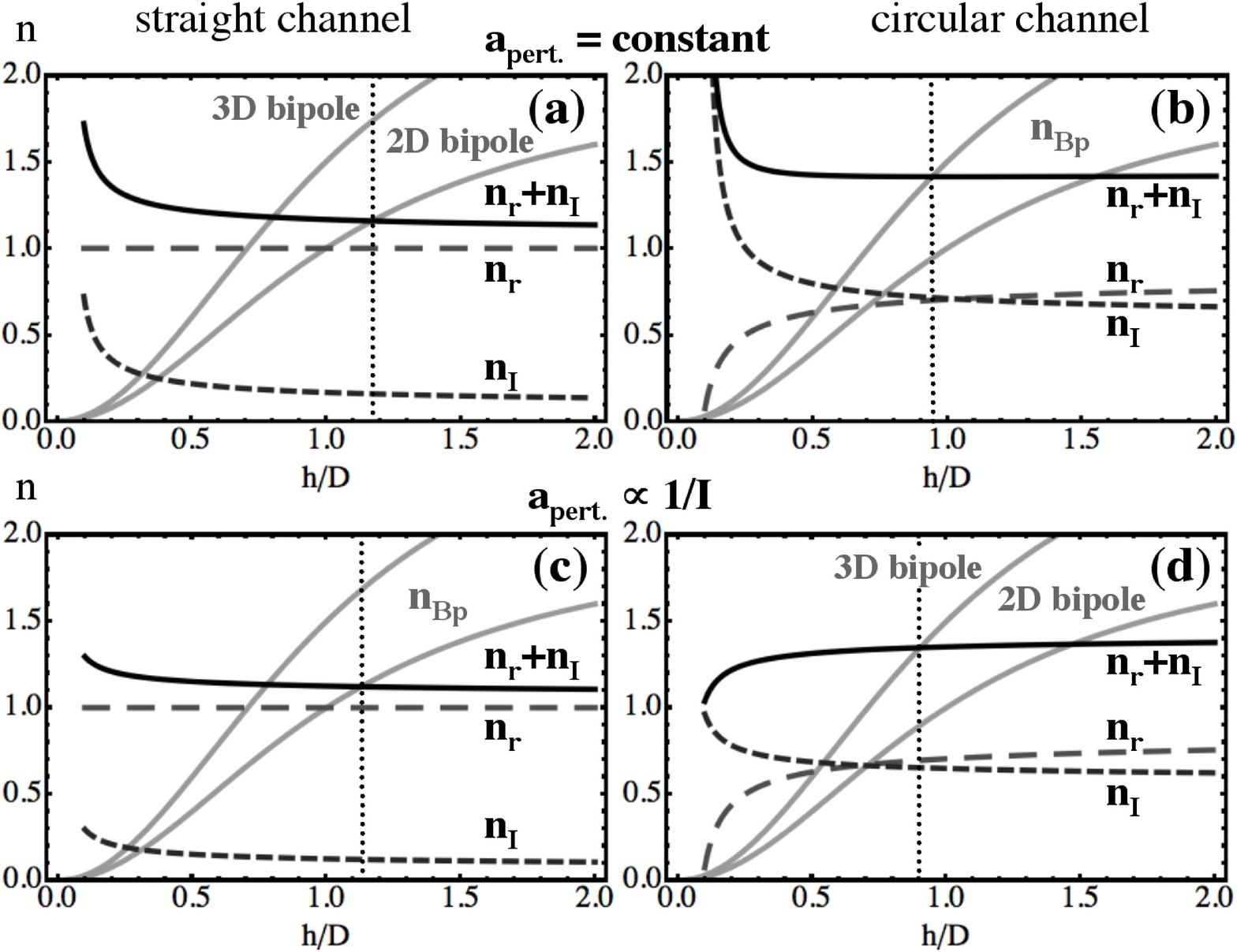}}
}{
  \centerline{\includegraphics[width=0.48\textwidth, clip=]{f3_dlog_terms}}
} 
\caption{ Dependence of the decay indexes 
$\nr=-\partial \ln r / \partial \ln h$, 
$\nI=-\partial \ln |I| / \partial \ln h$ and
$\nBp=-\partial \ln |\Bpx| / \partial \ln h$ for a straight (resp. circular) current channel in the left (resp. right) column.  The critical index $\nBpcrit = \nr + \nI$ for instability is shown with a black continuous line. The decay indexes of 2D and 3D potential bipoles are shown in both columns for reference. The dotted vertical line indicates the critical height (point ``c'' in Figure~\ref{Fig_non_equil}).  All panels are drawn with the normalized equilibrium radius $a/D=0.1$ and for $h>a$. The top (resp. bottom) panels have $\na=-\partial \ln a / \partial \ln |I| =0$ (resp. $\na=1$).
}
 \label{Fig_dlog_terms}
\end{figure}

%%%%%%%%%%%%%%%%%%%%%%%%%%%%%%%%%%%%%%%%%%%%%%%%%%%%%%%%%%%%%%%%%%%%%%%%%%
\begin{figure}[t!]    
\IfFileExists{submit.txt}{
  \centerline{\includegraphics[width=0.8\textwidth, clip=]{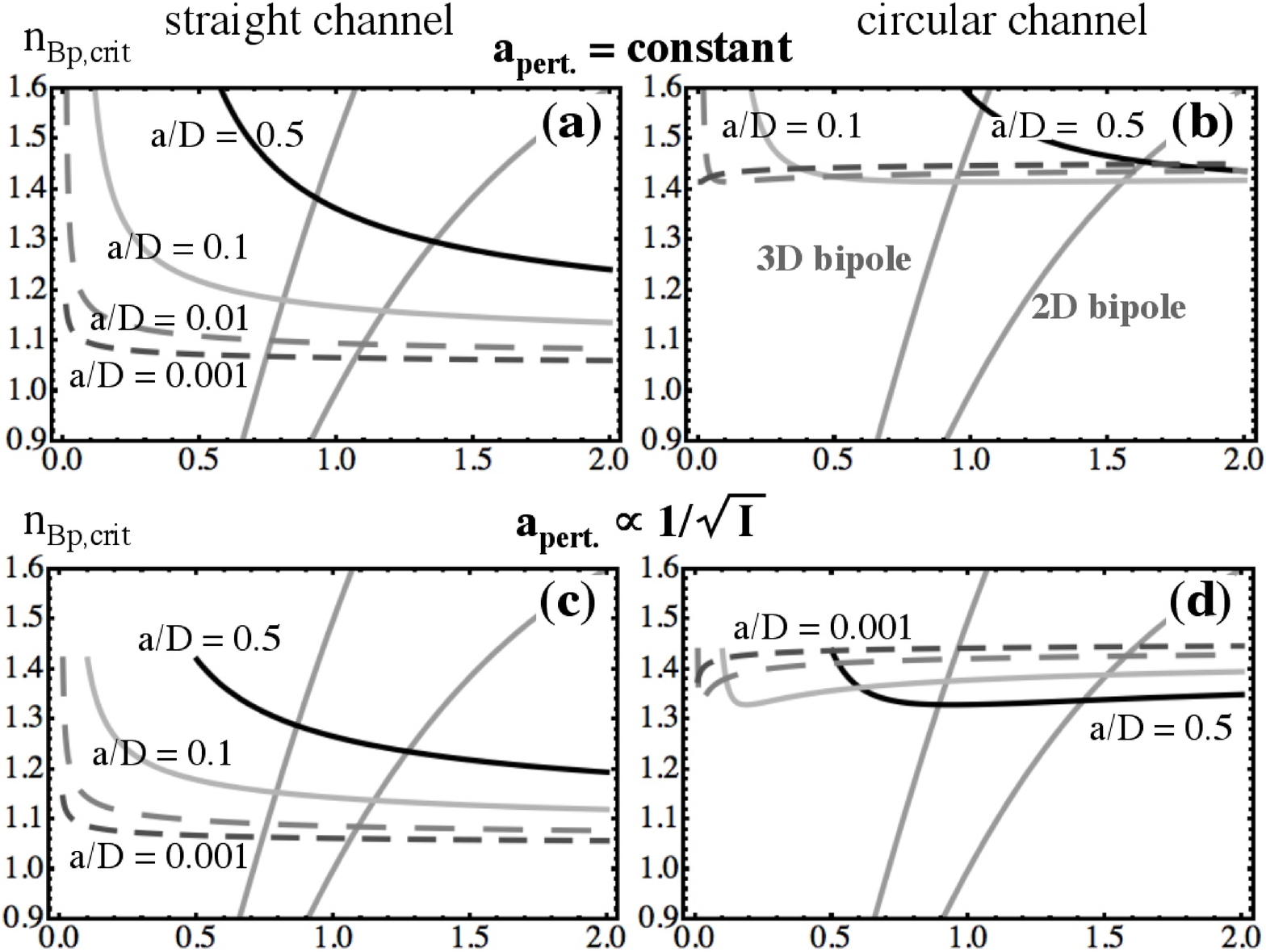}}
  \centerline{\includegraphics[width=0.8\textwidth, clip=]{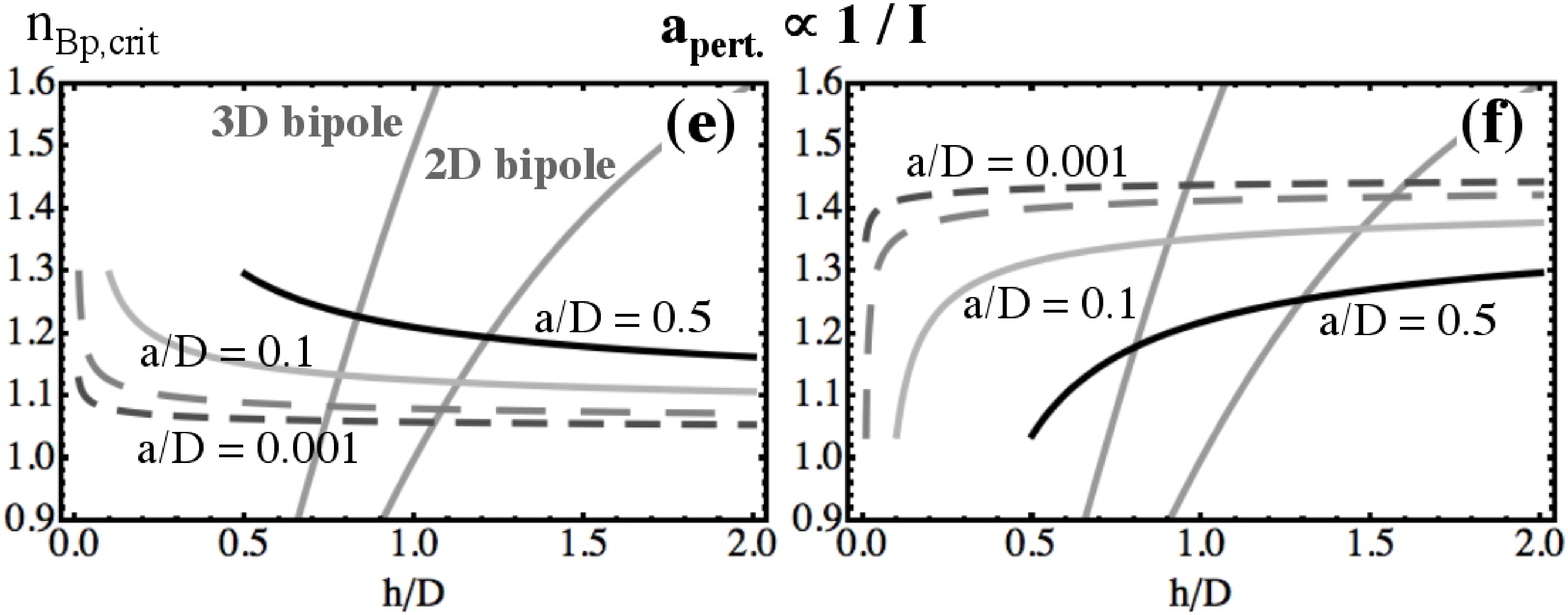}}
}{
  \centerline{\includegraphics[width=0.48\textwidth, clip=]{f4_nBcritA}}
  \centerline{\includegraphics[width=0.48\textwidth, clip=]{f4_nBcritB}}
} 
\caption{The critical index $\nBpcrit = \nr + \nI$ for instability with various equilibrium radius $a$ normalized by $D$ (half distance between photospheric field concentrations). The results are for a straight (resp. circular) current channel in the left (resp. right) column. The flux rope expands more during the perturbation (larger $\na$ value) from top
($\na=0$) to bottom ($\na=1$). The $\nBpcrit$ curves are drawn for $h>a$. The decay indexes, $\nBp$ of 2D and 3D potential bipoles are shown in both columns for reference.
}
 \label{Fig_nBcrit}
\end{figure}

\subsection{Example of a Straight Current Channel} %%%%%%%%%%%%
   \label{Stab.4}

   %{\S\bf Contributions to $I$ change with flux conservation}\\
  We follow the same derivation than in previous subsection but for a current channel formed by two parallel lines (Figure~\ref{Fig_schema}(c)). In this geometry, the external inductance $\Le$ is given by Equation~(\ref{L_s}), again with $\li =0$.  The repulsion $r$ is given by Equation~(\ref{rep_s}) and $P=2\Delta y$ (with $\Delta y \gg h,D$).  For $\na=0$, the contribution from the field created by the current channel to the decay index $\nI$, Equation~(\ref{n_I}), is exactly twice the contribution from the potential field, but with opposite sign (as above for the circular channel, with $\lf =0$)
  \begin{equation} \label{dlogI_s}
  \frac{\partial \ln \Le}{\partial \ln h} = 2\, \frac{P\,h\,\rep}{\Le} 
                                          = \frac{1}{\ln (2h/a)}    \,.
  \end{equation}

   %{\S\bf Contributions to $I$ change with flux conservation}\\
With Equation~(\ref{rep_s}), the decay index of the repulsion $r$ is simply $n_r =1$.
Combining the above results, the instability condition, Equation~(\ref{unstable}), for a straight current channel is  
  \begin{equation}  \label{unstable_s}
  \nBp > 1  + \frac{1}{2(\ln (2h/a) + \na)}      \,.
  \end{equation}
In the limit of a very thin current channel ($a \ll h$), the instability threshold is close to 1, as found by \citet{vanTend78}.

\subsection{Comparaison of Circular and Straight Current Channels} %%%%%%%%%%%%
   \label{Stab.5}

   %{\S\bf $\nI$ contribution to the stability}\\
From Equation~(\ref{unstable}), the critical decay index of the potential field for instability, $\nBpcrit$ has two contributions: the decay indices of the repulsion, $\nr$, and the decay index of the current during the ideal perturbation, $\nI$.  The main difference of stability between the above two current channels is a much lower index $\nI$ for a straight channel (Figure~\ref{Fig_dlog_terms}).   This is due to the low dependance of $L_{\rm e}$ on the height $h$ (compare Equation~(\ref{L_s}) to Equation~(\ref{L_c})).    In the limit $a \ll h$, this is the origin of a more stable circular current channel ($\nBpcrit \approx 3/2$) than a straight current channel ($\nBpcrit\approx 1$). However, this limit is not applicable to the eruptive coronal configurations since it requires extremely thin current channels: for example, even with $a/D=10^{-3}$, this limit is only weakly approached (Figure~\ref{Fig_nBcrit}).  This is due to the $\ln (h/a)$ dependance in both Equations~(\ref{unstable_c}) and (\ref{unstable_s}).

   %{\S\bf $\nr$ contribution to the stability}\\
Moreover, the above difference is partly compensated by a lower decay index of the repulsion, $\nr$, for a circular channel (as the repulsion, $r(h)$, is decreasing slower with height, especially for large values of $a$).  The net result is that the critical decay index, $\nBpcrit$, for circular and straight channels, with finite radius $a$, are much closer than in the limit $a \ll h$. This is illustrated in Figure~\ref{Fig_dlog_terms} with $a/D=0.1$, which still corresponds to a relatively narrow channel.  This effect is amplified with a flux rope having a larger expansion during the perturbation, so a larger $\na$ value, because the contribution of $\nI$ to $\nBpcrit$ is reduced.  Indeed, even with a relatively thin flux rope, $a/D=0.1$, and an internal expansion rate $\na=1$, as in \citet[][]{Lin98}, $\nBpcrit$ values are close for circular and straight current channels: $\approx 1.3$ and $\approx 1.1$, respectively (Figure~\ref{Fig_dlog_terms}(c) and (d)).  They correspond to a comparable critical height ($h/D \approx 1$) for straight and circular channels with a 2D and 3D bipole (Equations~(\ref{bx_c}) and (\ref{bx_s}), respectively).  

   %{\S\bf Effect of the flux rope radius}\\
In previous studies, the instability threshold $\nBpcrit$ was typically taken in the limit of very thin current channels \citep[e.g.][]{vanTend78,Kliem06}. The above analytical theory is indeed done in the limit of thin current channels (typically $a/D \leq 0.1$).  In fact, relatively broad current channels are expected in the coronal with magnetic extrapolations \citep[e.g.][]{Schrijver08,Savcheva09} and are present in MHD simulations \citep[e.g.][]{Fan04,Torok07,Aulanier10}. So, we also show the approximative results for a broad channel case, $a/D=0.5$.
For a straight current channel, the flux conservation provides an increasing stabilizing effect as the channel radius, $a$, is increasing (Figure~\ref{Fig_nBcrit}). Indeed, for a broad \p{straight} channel ($a/D \geq 0.1$), $\nBpcrit$ can reach a value comparable, or even larger in some casesthan the one obtained for the corresponding circular channel with the same parameter values \p{(Figure~\ref{Fig_nBcrit}(e) and (f))}.  We conclude that the circular and straight current channels have typically comparable instability threshold values for the range of parameters expected in the corona. 

   %{\S\bf Effect of the internal evolution}\\
The internal evolution of the current channel during the perturbation (i.e. the effect of $\na$), has a {\bf common} effect on the stability of a circular and of a straight current channel \p{because the stabilizing term provided by the flux conservation is generally decreased with an increase of $\na$
(Figures~\ref{Fig_dlog_terms},\ref{Fig_nBcrit}). However, changing the channel thickness ($a/D$) has an opposite effect on the instability threshold for a circular and for a straight current channel}
(compare the panels in Figure~\ref{Fig_nBcrit} where the $a/D$ curved are ordered oppositely in the two columns).  For a straight channel, $\nr =1$, so $\nBpcrit=\nr + \nI$ is affected only by the dependance of $\nI$ on $a/D$.  But, for a circular channel, the $\nr$ dependance on $a/D$ is important, and dominates the contribution of $\nI$, as soon as $\na$ is slightly positive.   It implies a dependance of $\nBpcrit$ on $a/D$ for a circular channel that is opposite to that for a straight channel.      

%The internal evolution of the current channel during the perturbation (i.e. the effect of $\na$), has an opposite effect on the stability of a circular and of a straight current channel (compare the panels in Figure~\ref{Fig_nBcrit} where the $a/D$ curved are ordered oppositely in the two columns).  In fact, there is a common effect of $\na$ on the stability: as $\na$ is increased, the stabilization provided by the flux conservation is generally decreased, as shown by a lower decay index $\nI$ (Equation~(\ref{n_I})).    For a straight channel, $\nr =1$, so $\nBpcrit=\nr + \nI$ is affected only by the dependance of $\nI$ on $a/D$.  But, for a circular channel, the $\nr$ dependance on $a/D$ is important, and dominates the contribution of $\nI$, as soon as $\na$ is slightly positive.   It implies a dependance of $\nBpcrit$ on $a/D$ for a circular channel that is opposite to that for a straight channel.      

%%%%%%%%%%%%%%%%%%%%%%%%%%%%%%%%%%%%%%%%%%%%%%%%%%%%%%%%%%%%%%%%%%%%%%%%%%%
\section{Conclusion}   \label{Conclusion}
 
   %{\S\bf Summary of the problem}\\
How to destabilize a coronal magnetic configuration is a key-issue of CME research.  Among several possibilities, two candidates are a loss of equilibrium and a torus instability occuring during the evolution of the magnetic configuration.   Both have been initially developed with the approximation that the coronal currents are restricted to a non-neutralized current channel, and both theory were further analyzed with MHD simulations, relaxing part of the initial approximations of the analytical developments, but at the expense of not covering the parameter space.  
     
   %{\S\bf Difference loss of equilibrium / instability ?}\\
In this study, we revisit both analytical theories and compare their approaches for the two simple configurations where their results apparently differ: a straight and a circular current channel.   A loss of equilibrium is typically, but not always, present in both configurations when an ideal-MHD evolution is imposed during the long-term evolution of the magnetic configuration.  However, when a loss of equilibrium occurs, the magnetic configuration is also ideally unstable.  From the results of Sections~\ref{Loss} and \ref{Stab}, we conclude that both approaches are in fact both compatible and complementary.  In particular they agree on the position of instability, if no significant current sheets are formed during the long-term evolution of the magnetic configuration.  Moreover, slow resistive processes, e.g. tether cutting, are probably occurring all the way before an eruption occurs.  Therefore, we conclude that the analytical theory is most useful in deriving an instability threshold with the constraint of ideal MHD evolution on a short time scale (coronal Alfv\'en time) for a given magnetic equilibrium.  
%This stability analysis needs to be embedded in the context of a long-term evolution, which is more difficult to model analytically.   
   
   %{\S\bf Difference straight / circular ?}\\
We also compare the physical origin of the instability of straight and circular current channels.  In order to model the negligible evolution of the magnetic flux crossing the photosphere on the coronal Alfv\'en time scale, a theoretical image current is introduced below the photosphere
(Figure~\ref{Fig_schema}).  For a straight current channel, the repulsion of the image is balancing the Laplace force between the coronal current and the potential field (associated to the photospheric field distribution).  For a circular current channel, the repulsion of the nearby coronal current is also present (called hoop force).  However, since the repulsion force depends only on the global curvature radius and on the thickness of the current channel for a circular channel, it could lead the false conclusion that the repulsion force has a different origin for the straight and circular current channels.  In fact, as shown by \citet{Garren94}, both the coronal and the image currents generically contribute to the repulsion force of a current channel.   Both terms actually combine in a single expression for a circular channel, while there is no contribution of the coronal current for a strictly straight current channel.    The circular and straight current channels are simply two limits of the general case with specific properties.

   %{\S\bf Different instability threshold}\\
The instability occurs when the potential magnetic field decreases fast enough with height, more precisely when its decay index, $\nBp$ as defined by Equation~(\ref{nBp}), is larger than a critical value $\nBpcrit$.  At the limit of extremely thin current channels $\nBpcrit=1$ and $1.5$ for a straight and circular current channel, respectively.   In fact, we show that this difference is not due to the difference in the repulsion force, but that it is due to the constraint of ideal MHD (conservation of the coronal magnetic flux below the current channel).  Moreover, with the sole contribution of repulsion force to the instability threshold (i.e. $\nI=0$ in Equation~(\ref{unstable})), first $\nBpcrit<1$ for a circular current channel while $\nBpcrit =1$ for a straight channel, and second $\nBpcrit$ approaches $1$ for both circular and straight channel as the channel becomes very thin.  This further indicates that there is no real difference in the origin of the repulsion force for a straight and circular current channels.
    
   %{\S\bf same physics and similar threshold}\\
We conclude that the same physics is involved in the instability of circular and straight current channels, and that they are just two particular limiting cases of more general current paths.  For the typical range of current-channel thickness expected in the coronal, and present in MHD simulations, and for a current channel expanding during an upward perturbation, $\nBpcrit$ has close values for both circular and straight current channels (in the range [1.1,1.3], Figure~\ref{Fig_nBcrit}(e) and (f)). If the current channel would not expand, the decay index $\nBpcrit$ would be higher, typically in the range [1.2,1.5], but still not so different in both cases  (Figure~\ref{Fig_nBcrit}(a) and (b)).  Similar critical indexes have been found in MHD simulations starting from a initial equilibrium having a coronal current channel close to half torus \citep{Torok07,Schrijver08}.  Otherwise, from the measurement of the height of
a set of quiescent prominences, combined with potential field extrapolations, \citet{Filippov01} found $\nBpcrit \approx 1$.  This threshold is closer to $\nBpcrit$ of the straight current channel as expected since quiescent prominences are horizontally extended structures. 

   %{\S\bf Difference with simulation, why ?}\\
In an MHD simulation, where a flux rope and its associate current channel is progressively formed due to photospheric motions, flux cancelation, and magnetic reconnection, \citet{Aulanier10} found an unstable configuration when $\nBpcrit \approx 1.5$. There, the flux rope height satisfied $h/D \approx 1$.
This was in favor of the ``torus instability''.  However, with the above results, this
threshold would require that the current channel is almost rigid during the perturbation (i.e. $\na =0$ as in Figure~\ref{Fig_nBcrit}(b)).   \p{It is not obvious that this condition is met in} a low-$\beta$ magnetic field.  More relevant is probably the role of the anchorage of the current channel at fixed photospheric positions during the stability analysis, a constraint not present for both the straight and circular models studied above, but included in the non-equilibrium study of \citet{Isenberg07}.   The present analytical theory is also over simplifying the coronal current distribution to only one current channel, while at least partial current neutralization  as well as other current layers are typically present in MHD simulations.  This may also \p{raise} the critical index, $\nBpcrit$, to larger values \citep[e.g. $\nBpcrit$ as high as $1.9$ was found in the MHD simulations of flux emergence by][]{Fan07}.  The precise understanding of the instability threshold is important for determining when a CME would occur.  This will be the object of further developments of the analytical theory.
%, which are outside the scope of present paper.

\pc{
It would also be desirable to derive the critical index from observations of 
eruptive prominences and sigmoids, the pre-eruptive altitudes of which can either 
be measured using two vantage points (e.g. using the pair of STEREO imagers) or when 
they cross the solar limb. At first approximation, the background coronal magnetic 
field would then have to be extrapolated in the potential field approximation, 
using photospheric magnetograms either taken on the same day of the eruption 
if possible, or a few days before if no magnetogram is available. Such a survey 
of various eruptive solar features would extend the work carried out by \citet{Filippov01}, 
who focused on long and high altitude quiescent prominences, that mostly concern 
the straight channel model. 
}
\acknowledgments

The authors thank the referee for helpful comments which improved the clarity of the paper.
Financial supports by the European Commission through the FP6 SOLAIRE Network (MTRN-CT- 2006-035484)
and through the FP7 SOTERIA project (Grant Agreement no 218816)
are acknowledged.

%The research leading to these results has received
%funding from the European Commission's Seventh
%Framework Programme (FP7/2007-2013) under the grant
%agreement n¡ 218816 (SOTERIA project, www.soteria-space.eu).
%Financial support by the European Comission through
%the SOLAIRE network (MTRM-CT-2006-035484) is also
%gratefully acknowledged.

\bibliographystyle{apj}
\bibliography{stability}

\end{document}